\documentstyle[12pt,aaspp]{article}

\def\db{\frac{\partial}{\partial b}}
\def\lb{\left(}
\def\rb{\right)}
\def\lsb{\left[}
\def\rsb{\right]}
\def\ap{a^{'}_{1}}

\def\pl{\partial}

\def\xio{\xi^{(1)}(x,t)}
\def\ini{\int_0^{\infty}}

\def\xiox{\xi^{(1)} (x)}
\def\xioy{\xi^{(1)} (y)}
\def\xibox{\overline{\xi^{(1)}} (x)}

\def\txy{\theta_{xy}}
\def\xioxt{\xi^{(1)} (x,t)}
\def\xioyt{\xi^{(1)} (y,t)}
\def\xiboxt{\overline{\xi^{(1)}} (x,t)}
\def\xiboyt{\overline{\xi^{(1)}} (y,t)}
\def\txy{\theta_{xy}}
\def\ap{a^{'}}
\begin{document}

\title{The evolution of correlation functions in the Zel'dovich
approximation and its implications for the validity of perturbation
theory.} 
\author{Somnath Bharadwaj}
\affil{ Raman Research Institute, Bangalore 560 080, India.
 \altaffilmark{1} \altaffiltext{1}{Postal address} 
 \\ and \\ Joint Astronomy Program, Indian Institute of Science,
 Bangalore 560 012, India. \\  email somnath@rri.ernet.in} 
\authoraddr{Raman Research Institue Bangalore 560080 India}

\begin{abstract}
We investigate whether it is possible to study perturbatively the
transition  in cosmological clustering between a single streamed flow
to a multi streamed flow. We do this by considereing a system whose
dynamics is governed by the Zel'dovich approximation (ZA) and
calculating the evolution 
of the two point correlation function using two methods: 1. Distribution
functions 2. Hydrodynamic equations without pressure and vorticity. The
latter method breaks down once multistreaming occurs whereas the former
does not. We find that the two methods give the same results to all
orders in a perturbative 
expansion of the two point correlation function. We thus conclude that we 
cannot study the transition from a single stream flow to a
multi-stream flow in a perturbative  
expansion. We expect this conclusion to hold even if we use the full
gravitational dynamics (GD) instead of ZA. 

We use ZA to look at the evolution of the two point correlation
function at large spatial separations and we find that until the onset of
multi-streaming the evolution can be described by a diffusion process
where the linear evolution at large scales gets modified by the
rearrangement of matter on small scales. We compare these results
with the lowest order nonlinear results from GD.
We find that the difference is only in the numerical value of the
diffusion coefficient and
we interpret this physically. 

We also use ZA to study the induced three point correlation function.
At the lowest order of nonlinearity we find that, as in the case of
GD, the three point correlation does 
not necessarily have the hierarchical form. We also find that at large
separations the effect of the higher order terms for the three point
correlatin function is very similar to that for the the two point
correlation and 
in this case too the evolution can be be described in terms of a
diffusion process. 
\end{abstract}


\newpage

\section{Introduction}
The inviscid hydrodynamic equations without pressure and vorticity
(referred to as 
the HD equations in the rest of this paper) are often used to describe
the evolution of disturbances in an expanding universe filled with
collisionless particles that  interact only through Newtonian gravity.
The disturbances that are usually considered are such that initially all
the particles at any point have the same velocity i.e. it is a single
streamed flow. Such a situation is correctly described by the HD equations.
As the disturbances evolve the particle trajectories intersect and there
are particles with different velocities at  the same point i.e. the flow
becomes multi-streamed. When this occurs the HD equations are no longer
valid. This is because the HD equations neglect the local stress
tensor associated with the moments of the velocity about the mean
velocity at a point. 

The BBGKY hierarchy of equations obeyed by the distribution functions
can be used instead  of the HD equations. The distribution functions  
keep track of the position and velocity of the particles and these
equations are valid even if multistreaming occurs. 
   The question we would like to address in this paper is whether we can
study the effects of multistreaming by using distribution functions
perturbatively to follow the evolution of the disturbances. 
   
 We look at the perturbative evolution of the density - density two
point correlation function 
for Gaussian initial conditions in  a universe with $\Omega=1$.  The
perturbative expansion of this 
function using the HD equations has been studied by many authors
(Juskewicz 1981; Vishniac 1983; Fry 1994). 
 In a recent paper (paper II,Bharadwaj 1995) we have calculated the lowest
order nonlinear term for the two point correlation function using the
moments of the BBGKY hierarchy. These equations are based on the
distribution functions and are valid even in the multi-streamed regime.
The two different methods of calculation (HD and BBGKY) aro found to
give the same result  at the 
lowest order of nonlinearity, and hence, to this order, distribution 
functions have not been able to capture any effect of
multi-streaming. In this paper we investigate whether  by going to
higher orders of perturbation theory we shall be able to study any
effects of multi-streaming or if it is a limitation
of perturbation theory that it cannot follow the transition
from a single streamed flow to a multi streamed flow.  

Because of the difficulty in  calculating the higher order terms in a 
 perturbative treatment of gravitational dynamics (GD), we look at a simpler 
system where we use the Zel'dovich approximation  (ZA, Zel'dovich
1970) to determine the motion of the particles. In this situation too
the 
transition from a single streamed  flow to a multi-streamed flow occurs
and we can analyse it to see if in a perturbative calculation using
distribution functions we can include any effects of multi-streaming which
would be missed if the HD equations were used instead.
	
    In section 2 we discuss the evolution equations. In section 3 we use
distribution functions to calculate the evolution of the two point
correlation function. In section 4  we do the same calculation using the HD
equations and compare the result with that obtained in section 3.

Bond and Couchman (1988) have studied the evolution of the two point
correlation function using ZA and the calculation presented in section
3 is on similar lines. In a more recent paper Schneider and Bartlemann
(1995) have studied the evolution of the power spectrum in ZA. 
 For a comprehensive article on various aspects
of ZA the reader is referred to a review by Shandarin and Zel'dovich
(1989). 

 In paper II we investigated  the lowest order nonlinear 
correction (using GD) to the two point correlation for initial power
spectra of the form $P(k) \propto k^n$ at small $k$ and an exponential
or Gaussian cutoff at large $k$.  
 We found that for $0<n \le 3$, the  numerical results for the
nonlinear correction to the two point correlation function at  
large $x$ could be fitted by a simple formula . We also interpreted this
formula in terms of a simple diffusion process. In section 5 of this
paper we investigate the evolution of the two point correlation
function at large separations using ZA and compare it with the results
from GD.

   In section 6 we look at the evolution of the induced three point
correlation function using ZA. This was first calculated for GD by Fry
(1984) and he 
concluded that for power law initial conditions, at large separations,
the three point correlation function could be described by the
hierarchical form where it can be written in terms of products of the
two point correlation function evaluated at the separations involved. 
In an earlier paper (part I, Bharadwaj 1994) we calculated the same
quantity and found that these conclusions were not fully correct. We
showed that the three point correlation function at some length
scale, depends not only on the two point correlation at the same
length  scales but on all smaller scales also. As a result we found
that the hierarchical form is true for only a class of initial
conditions and that there was a class for which it did not hold. In
this paper we first calculate the three point correlation function at
the lowest order of nonlinearity for ZA and compare it to the results
from GD. We then go on to study the effect of the higher order
nonlinear terms at large separations.

The calculations using ZA are valid for any value of $\Omega_0$ but
whenever we make comparisons with GD it is for the specific value
$\Omega=1$. 

A similar calculation has been done by Grinstein and Wise (1987) who
have studied the evolution of skewness of the density field averaged
over a Gaussian ball. Also, Munshi and Starobinsky (1994) have
considered the evolution of the skewness of the density field for ZA 
and various other approximations, and
Bernardeau et. al. (1993) have calculated the evolution of the skewness of the
density field averaged over top hat filters. All of these calculations
have been done at the lowest order of nonlinearity.

In section 7 we present a discussion of the results obtained  and the
conclusions. 
    
\section{Evolution of the distribution fuction} 
The Zel'dovich approximation defines a map from the initial position af a particle
to its position at any later instant. If $x_{\mu}(t)$ is the comoving
coordinate of a particle at any time $t$,  the initial instant being
$t_0$,  and  $b(t)$ the growing mode in the linear analysis of density
perturbations, this map is
\begin{equation}
x_{\mu}(t)=x_{\mu}(t_0) + b(t) u_{\mu} \,. \label{eq:a1}
\end{equation}

The quantity $u_{\mu}$ is related to the peculiar velocity
$\rm{v}_{\mu}(t)$  
at any instant  by 
\begin{equation}
\rm{v}_{\mu}(t) = a(t)  \frac{d}{d t} x_{\mu}(t)=a(t) \dot{b}(t)  u_{\mu}
\end{equation}
where $a(t)$ is the scale factor.

We consider a system of particles whose motion is governed by this
mapping. 
This  can be described by a distribution function $f(x,u,t)$,
where  $f(x,u,t) d^{3}x d^{3}u$ is the number of particles in the
volume  $d^  
{3}x$ around the point $x$ and having a value of $u$ in an interval
$d^{3}u$ around $u$.

 We can see that Liouville theorem is true for the mapping defined in 
 equation (\ref{eq:a1}). Using this we can obtain the equation 
 for the time evolution of the
distribution function $f$,
\begin{equation}
f(x,u,t)=f(x- b(t) u ,u,t_{0}) \,. \label{eq:a2}
\end{equation}

We can also use equation (\ref{eq:a1}) to obtain a
differential equation for the evolution of the distribution function
\begin{equation}
\db f(x,u,b) + u_{\mu} \frac{\pl}{\pl x_{\mu}} f(x,u,b)=0 \,, \label{eq:a3}
\end{equation}
where we use the growing mode $b$ instead of time as the evolution parameter.

We are interested in the evolution of the statistical properties of an
ensemble of such systems.

 Every member of the ensemble initially has the particles uniformly
distributed. Initially each particle can be labeled by its coordinate
$x_{\mu}$. The
 particles are given velocities $u_{\mu}(x)$. The velocity
field is the
gradient of a function  $\psi(x)$ which for each system is a different
realization of a Gaussian random field. It is assumed
that $\psi$ is statistically homogenous and isotropic. The statistical
 properties of the
ensemble are initially fully specified by the two point correlation of
$\psi$ which is defined as $\phi(x)=<\psi(0)\psi(x)>$, where the angular 
brackets $<\, > $ denote ensemble averageing.      

The statistical quanitity whose evolution we shall focus on in this paper
is the density two point correlation function $\xi(x,t)$ . This is defined
 by the relation 
\begin{equation}
<\rho>^2 (1+\xi(x))=<\rho(0)\rho(x)> \,.
\end{equation}
where $\rho(x)$ is the mass density. This is just the number
density of particles multiplied by the mass
of each particle which  is assumed to be the same for all the particles.

\section{The two point correlation using distribution functions}
In this section we look at the evolution of the ensemble averaged
two point distribution functions  $\rho_{2}$.
This is defined as
\begin{equation}
\rho_{2}(x^{1},x^{2},u^{1},u^{2},t)=<f(x^{1},u^{1},t) f(x^{2},u^{2},t) >.
 \end{equation}
 From homogeneity and isotropy we can also say that 
 \begin{equation}
  \rho_{2}(x^{1},x^{2},u^{1},u^{2},t)=\rho(x,u^{1},u^{2},t)
  \end{equation}
  where 
  \begin{equation}
x_{\mu}= x^{2}_{\mu}-x^{1}_{\mu}.
  \end{equation}
The density two point correlation function is related to  the zeroth moment
 of the two point distribution function with respect to $u$.
 
\begin{equation}
<\rho>^2(1+\xi(x,t))=\int \rho_2 (x,u^{1},u^{2},t) d^{3}u^{1}d^{3}u^{2}.
\label{eq:t1}
\end{equation}

In this paper we normalize $<\rho>=1$.

The initial two point distribution is a Gaussian in the velocities and
hence specified by the covariance matrix
\begin{equation}
T^{a b}_{\mu \nu}(x)=<u^{a}_{\mu} u^{b}_{\nu}>(x)=\int u^{a}_{\mu} u^{b}_{\nu} 
\rho_{2}(x,u^{1},u^{2}, t_{0}) d^{3} u^{1}  d^{3} u^{2} 
\end{equation}
where $a,b$ take values $1,2$. 
The initial two point distribution function then is the Gaussian distribution  
\begin{equation}
\rho_{2}(x,u^{1},u^{2},t_{0})= \frac{1}{ \lb2 \pi \rb^3 \sqrt{ \Delta T(x)}}
  \exp \lsb -\frac{1}{2} u^{a}_{\mu} u^{a}_{\mu} (T^{-1})^{a b}_{\mu \nu}(x)
  \rsb \,,   \label{eq:a6}
\end{equation}
where $\Delta T(x)$ is the determinant of the covariance matrix.
In terms of the potential  $\phi$ we have 
\begin{equation}
<u^{1}_{\mu} u^{2}_{\nu}>=-\pl_{\mu}  \pl_{\mu} \phi(x)
\end{equation}
and 
\begin{equation}
<u^{1}_{\mu} u^{1}_{\nu}>=- \frac{1}{3}\nabla^{2} \phi(0) \delta_{\mu \nu}.
\end{equation}

  We use equation (\ref{eq:a2})  to obtain the time evolution of $\rho_{2}$ 
  \begin{equation}
 \rho_{2}(x,u^{1},u^{2},t)=\rho_{2}(x-(u^{2}-u^{1}) b(t),u^{1},u^{2},t_{0}).
 \label{eq:a4}
 \end{equation} 

This may also be written as 
\begin{equation}
\rho(x,u^{1},u^{2}, t)=\int \delta^{3}\lsb x^{'}-\lb x-(u^{2}-u^{1}) b\lb t\rb \rb \rsb 
\rho_2 (x^{'},u^{1},u^{2}, t_{0}) d^{3}x^{'}\,.
\end{equation}
Using the Fourier expansion of the Dirac delta function and using equation 
(\ref{eq:a6})  we have
\begin{eqnarray}
\rho(x,u^{1},u^{2}, t)&=&\int \lb \frac{1}{2 \pi} \rb ^{3} \exp \lsb
i k_{\mu}  \lb x^{'}_{\mu}- x_{\mu} \rb \rsb  \exp \lsb i k_{\mu} 
\lb u^{2}_{\mu}-u^{1}_{\mu} \rb  b\lb t\rb  \rsb  
\nonumber \\ &\times &
 \frac{1}{ \lb2 \pi \rb^3 \sqrt{ \Delta T(x^{'})}}
 \exp \lsb -\frac{1}{2} u^{a}_{\mu} u^{a}_{\mu} 
(T^{-1})^{ab}_{\mu \nu}(x^{'}) \rsb d^{3}k d^{3}x^{'}\label{eq:a7}
\end{eqnarray}

Using this in equation (\ref{eq:t1}) and doing the $u$ integrals   we get 
\begin{equation}
1+\xi(x,t)=\lb \frac{1}{2 \pi} \rb^{3} \int \exp \lsb i k_{\mu} 
\lb x^{'}_{\mu}- x_{\mu} \rb \rsb \exp \lsb -\frac{b^{2}(t)}{2}
k_{\mu}k_{\nu} F_{\mu \nu}(x^{'}) \rsb d^{3}x^{'} d^{3}k \,,
\label{eq:a8} 
\end{equation}
where 
\begin{equation}
F(x)_{\mu \nu} = -\frac{2}{3}\nabla^{2} \phi(0) \delta_{\mu \nu}+ 2 \pl_{\mu}
\pl_{\mu} \phi(x)
\end{equation}
Doing the $k$ integral we obtain the two point correlation as 
\begin{equation}
1+\xi(x,t)= \frac{1}{\lb 2 \pi \rb^{\frac{3}{2}} b^3(t)} \int \frac{1}
{\sqrt {\Delta F(x^{'})}} \exp \lsb - \frac{1}{2 b^{2}(t)}\lb
x^{'}_{\mu}-x_{\mu} 
 \rb \lb x^{'}_{\nu}-x_{\nu} \rb F^{-1}_{\mu \nu}(x^{'}) \rsb d^{3}x^{'}\,.
 \label{eq:a9}
\end{equation}
 Instead of integrating equation (\ref{eq:a8}), if we do a Taylor
expansion of
\begin{eqnarray}
 exp\lsb -\frac{b^{2}(t)}{2}k_{\mu}k_{\nu} F_{\mu \nu}(x^{'}) \rsb
\nonumber 
\end{eqnarray}
 and then do the $k$ and the $x'$ integrals, we obtain 

\begin{eqnarray}
1+ \xi(x,t)&=&\sum_{n=0}^{\infty} \frac{b^{2n}}{n!}
\pl_{\mu_1} \pl_{\nu_1}... \pl_{\mu_n} \pl_{\mu_n}
\lsb \lb \pl_{\mu_1}\pl_{\nu_1} \phi(x)-\delta_{\mu_1 \nu_1}  
\frac{\nabla^2 \phi(0) }{3} \rb ... \right. \nonumber \\  
& & \left. ...
\lb \pl_{\mu_n}\pl_{\nu_n} \phi(x)-\delta_{\mu_n \nu_n} 
\frac{\nabla^2 \phi(0) }{3} \rb \rsb \label{eq:a10} \,.
\end{eqnarray}

Nowhere above has any assumption been made about the number of streams in
the flow . Equation (\ref{eq:a9}) obviously has the effects of
multistreaming built into it. Equation (\ref{eq:a10}) is what one would
obtain if one did a perturbative expansion of the distribution function
and calculated the two point correlation function. Whether by doing
the perturbative analysis this way (i.e. using distribution functions) we are able  to include the effects of
multistreaming is what has to be checked. 

\section{The two point correlation using the  hydrodynamic equations}
In this section we shall work in the single stream approximation. We
cosider any one member of the ensemble described previously. Its
evoulution is described by equation (\ref{eq:a3}). If we take the zeroth
moment of this equation with respect to $u$. Using the definitions 
\begin{equation}
\rho(x,b)=m \int f(x,u,b) d^{3} u
\end{equation}
and 
\begin{equation}
\rho(x,b) v_{\mu}(x,b)=m \int u_{\mu} f(x,u,b) d^{3} u
\end{equation}
 we have the continuity equation
\begin{equation}
\db \rho(x,b)+\pl_{\mu} ( \rho(x,b) v_{\mu}(x,b))=0\,.\label{eq:b1}
\end{equation}
Next, taking the first moment of equation (\ref{eq:a3}) and using equation
(\ref{eq:b1}) we have 
\begin{eqnarray}
& &\rho(x,b)[\db v_{\mu}(x,b) + v_{\nu}(x,b) \pl_{\nu} v_{\mu}(x,b) ] +
\nonumber \\ &+& 
m \pl_{\nu} \int (v_{\nu}(x,b)-u_{\nu}) ( v_{\mu}(x,b)-u_{\mu}) f(x,u,b) d^3 u
=0 \,. \label{eq:b2}
 \end{eqnarray}
 In the single stream approximation the last term in the above equation is
dropped, and we have 
\begin{equation}
\db v_{\mu}(x,b) + v_{\nu}(x,b) \pl_{\nu} v_{\mu}(x,b)=0\,. \label{eq:b3}
\end{equation}
We shall use use equations (\ref{eq:b1}) and (\ref{eq:b3}) to
perturbatively evolve the density and velocity fields of the system.
We then take ensemble averages and 
use these equations  to calculate the two point correlation function.

Using equation (\ref{eq:b1}) we can obtain an equation for the first
derivative of the two point correlation function 
\begin{equation}
\db [<\rho>^2 (1+\xi(x,b))]=-<\pl^1_{\mu}(\rho(x^1) v_{\mu}(x^1)) \rho(x^2)>
-<\rho(x^1) \pl^2_{\mu}(\rho(x^2) v_{\mu}(x^2)) > \,.
\end{equation}
Using the normalization $<\rho>=1$, the above equation may be written as
\begin{equation}
\db \xi(x,b)=-\pl^{a_1}_{\mu_1}<\rho(1) v^{a_1}_{\mu_1} 
\rho(2)>\,.
\end{equation}
We can use equation (\ref{eq:b1}) and (\ref{eq:b3}) to obtain equations
for the higher derivatives of the two point correlation
\begin{equation}
\frac{\pl^n}{\pl b^n} \xi(x,b))=(-1)^n \pl^{a_1}_{\mu_1}
\pl^{a_2}_{\mu_2}... \pl^{a_n}_{\mu_n}<\rho(1) v_{\mu_1}^{a_1}
v^{a_2}_{\mu_2}... v^{a_n}_{\mu_n} \rho(2)>\,.\label{eq:b4}
\end{equation}
Next we write   the two point correlation function as a Taylor series in
powers of the growing mode $b$
\begin{equation}
\xi(x,b)=\sum_{n=1}^{\infty}\frac{b^n}{n!} \frac{\pl^n}{\pl b^n} \xi(x,b)_{b=0}\,.
\end{equation}
It should be noted that this allows us to express the two point
correlation function at any instant in terms of the derivatives of the
two point correlation function at the initial instant.
Next,  using equation (\ref{eq:b4}) we get 
\begin{equation}
\xi(x,b)= \sum_{n=1}^{\infty}\frac{b^n(-1)^n}{n!} \pl^{a_1}_{\mu_1}
\pl^{a_2}_{\mu_2}... \pl^{a_n}_{\mu_n}<\rho(1)v^{a_1}_{\mu_1}
v^{a_2}_{\mu_2}... v^{a_n}_{\mu_n} \rho(2)>_{b=0}\,.\label{eq:b5}\,.
\end{equation}
Then using the fact that the initial density is  uniform, we can write
the two point correlation function at any arbitrary time in terms of
the initial velocities only i.e.
\begin{equation}
\xi(x,b)= \sum_{n=1}^{\infty}\frac{b^n}{n!}(-1)^n \pl^{a_1}_{\mu_1}
\pl^{a_2}_{\mu_2}... \pl^{a_n}_{\mu_n}<v^{a_1}_{\mu_1}
v^{a_2}_{\mu_2}...v^{a_n}_{\mu_n} >_{b=0}\,.\label{eq:b6}\,.
\end{equation}
Also the initial velocity field is Gaussian and hence all the odd terms in
equation (\ref{eq:b6}) are zero. We can then write this equation as 
\begin{equation} 
\xi(x,b)= \sum_{n=1}^{\infty}\frac{b^{2n}}{(2n)!} \pl^{a_1}_{\mu_1}
\pl^{b_1}_{\nu_1}...
\pl^{a_n}_{\mu_n} \pl^{b_n}_{\nu_n} <v^{a_1}_{\mu_1}
v^{b_1}_{\nu_1}... v^{a_n}_{\mu_n}  v^{b_n}_{\nu_n} >_{b=0}\,.\label{eq:b7}\,.
\end{equation}
 For the  Gaussian initial velocity field we have 
\begin{equation}
<v^{a_1}_{\mu_1} v^{b_1}_{\nu_1} v^{a_2}_{\mu_2} v^{b_2}_{\nu_2} ...
v^{a_n}_{\mu_n}  v^{b_n}_{\nu_n} > =\sum_P <v^{a_1}_{\mu_1} v^{b_1}_{\nu_1}>
<v^{a_2}_{\mu_2} v^{b_2}_{\nu_2} >...<v^{a_n}_{\mu_n} v^{b_n}_{\nu_n}>
\,,
\end{equation}
where the sum is over all possible ways of pairing the $u$'s.

Using this and the fact that the derivatives are symmetric in all the
indices involved, we have for the initial velocity field  
\begin{eqnarray}
& &\pl^{a_1}_{\mu_1} \pl^{b_1}_{\nu_1}
... \pl^{a_n}_{\mu_n} \pl^{b_n}_{\nu_n} 
<v^{a_1}_{\mu_1} v^{b_1}_{\nu_1}  ...
v^{a_n}_{\mu_n}  v^{b_n}_{\nu_n} > \nonumber \\ &=& \frac{(2n)!}{n! 2^n}
\pl^{a_1}_{\mu_1} \pl^{b_1}_{\nu_1} 
... \pl^{a_n}_{\mu_n} \pl^{b_n}_{\nu_n}\lsb
<v^{a_1}_{\mu_1} v^{b_1}_{\nu_1}>
...<v^{a_n}_{\mu_n} v^{b_n}_{\nu_n}>
\rsb \,.
\end{eqnarray}
This, when used in equation (\ref{eq:b7}), give us 
\begin{equation}
\xi(x,b)== \sum_{n=1}^{\infty} \frac{b^{2n}}{n! 2^n}
\pl^{a_1}_{\mu_1} \pl^{b_1}_{\nu_1} \pl^{a_2}_{\mu_2}\pl^{b_2}_{\nu_2}
... \pl^{a_n}_{\mu_n} \pl^{b_n}_{\nu_n}
\lsb <v^{a_1}_{\mu_1} v^{b_1}_{\nu_1}>
<v^{a_2}_{\mu_2} v^{b_2}_{\nu_2} >...<v^{a_n}_{\mu_n} v^{b_n}_{\nu_n}>\rsb
_{b=0}\,.
\end{equation}
Summing the superscripts $a_1,b_1,...a_n,b_n$ over the values $1$ and $2$
and using the fact that for the initial velocity field   
\begin{equation}
<v^a_{\mu}v^b_{\nu}>=\frac{-\nabla^2 \phi(0)}{3} \delta_{\mu \nu} \;
{\rm if}\,\, a=b
\end{equation}
and
\begin{equation}
<v^a_{\mu}v^b_{\nu}>= \pl^a_{\mu}\pl^b_{\nu} \phi(x) \; {\rm if} \,\,
a \neq b \end{equation}
we have 
\begin{eqnarray}
1+ \xi(x,t)&=&\sum_{n=0}^{\infty} \frac{b^{2n}}{n!}
\pl_{\mu_1} \pl_{\nu_1}... \pl_{\mu_n} \pl_{\mu_n}
\lsb \lb \pl_{\mu_1}\pl_{\nu_1} \phi(x)-\delta_{\mu_1 \nu_1}  
\frac{\nabla^2 \phi(0) }{3} \rb ... \right. \nonumber \\  
& & \left. ...
\lb \pl_{\mu_n}\pl_{\nu_n} \phi(x)-\delta_{\mu_n \nu_n} 
\frac{\nabla^2 \phi(0) }{3} \rb \rsb  \,.
\end{eqnarray}

This is the same as equation (\ref{eq:a10}) which was obtained using 
distribution functions. Thus we see that the perturbative calculation
of the two point correlation function using distribution functions has
no effects of multistreaming and hence we reach the conclusion that it
is not possible to perturbatively follow the transition from a single
streamed flow to a multi streamed flow. 

\section{The two point correlation at large separations.}
In this section we investigate the evolution of the two point
correlation function in the regime where it can be studied
perturbatively and we look at the behaviour at large separations. The
initial conditions for the evolution of the cosmological correlations
may be expressed in terms of the potential $\phi(x)$ or equivalently
in terms of the matter two point correlation in the linear epoch,
$\xio$. The two are related by the equation 
\begin{equation}
\xio=b^2  (t) \nabla^4 \phi(x) \,. \label{eq:w1}
\end{equation} 
Usually the initial conditions are given in terms of the matter two
point correlation $\xio$ or  its Fourier transform $b^2 (t) P_1 (k)$
which is the power spectrum. One then has to invert equation
(\ref{eq:w1}) to obtain the potential $\phi(x)$ and its derivatives.
In doing so one has the freedom in choosing boundary conditions and
the effect of changing the boundary condition is 
\begin{equation} 
\nabla^2 \phi(x) \rightarrow \nabla^2 \phi(x) + C_1
\end{equation}
and 
\begin{equation}
\phi(x) \rightarrow  \phi(x) + \frac{C_1 x^2}{6} + C_2 \,.
\end{equation}   
Equation (\ref{eq:a10})  for the two point correlation function is invariant uder these transformations and we are free to choose any boundary condition.
For initial conditions where the integral $\ini \xio x dx$ (or $\ini
P_1 (k) dk$) is finite we can choose the boundary condition ${\rm limit}_{x
\rightarrow \infty} \nabla^2 \phi(x)=0$. We then have  
\begin{equation}
<u^2>=-\nabla^2 \phi(0)= \ini \xiox x dx \label{eq:w2} \,.
\end{equation} 
  In addition, if at large $x$ the function  $\pl_{\mu} \pl_{\nu}
\phi(x)$ is monotonically decreasing and  $\pl_{\mu} \pl_{\nu} \phi(x)
\ll (\delta_{\mu \nu} /3) \nabla^2 \phi(0)$, we can then neglect
all but one of the $ \pl_{\mu} \pl_{\nu} \phi(x)$ terms that appear in
equation (\ref{eq:a10}).   For initial conditions where the power
spectrum has the form $P(k)\propto k^n$ at small $k$ and if it has a
cutoff at large $k$, the conditions discussed above are satisfied for
$n > -1$.  For these cases we obtain for the two point
correlation function at large $x$ 
\begin{equation}
\xi(x,t)=\sum_{n=0}^{\infty} \frac{b^{2(n+1)}}{n!}\lb \frac{-\nabla^2 \phi(0)}{3} \rb^n \lb \nabla^2 \rb^n \nabla^4 \phi(x) \label{eq:w3} \,.
\end{equation}

Using this we obtain the lowest order nonlinear correction to the two
point correlation function at large $x$,
\begin{equation}
\xi^{(2)} (x,t)=\frac{ b^2}{3} <u^2> \nabla^2 \xi^{(1)}(x,t)\label{eq:w4}
\end{equation}  
In paper II we have calculated the same quantity using GD and we found
that for $0<n \le 3$ at large $x$ the results can be fitted by the formula
\begin{equation}
\xi^{(2)} (x,t)=.194 b^2 <u^2> \nabla^2 \xi^{(1)}(x,t)\label{eq:w5}
\end{equation} 
We find that the two equations are very similar and they differ only in
the numerical coefficient. In paper II we also interpret equation
(\ref{eq:w5}) 
in terms of a simple heuristic model based on a diffusion process. We
consider  a particular
member of the enemble and look at the evolution of the density in volume
elements located at a separation $x$ from each other. 
 We assume that the density in each volume element grows according to
linear theory and the volume elements get rearranged randomly on small scales
because of their peculiar velocities. Based on this model we obtained an
equation identical to equation (\ref{eq:w4}). Thus we see that this
model gives an 
exact description of what happens in ZA at large
scales in the regime when the perturbative treatment is valid. In ZA, like
in our heuristic model, the velocity of the particles is fixed whereas in
GD the particle velocity changes as evolution proceeds. We believe
that this is responsible for the smaller diffusion coefficient for GD
compared to ZA. 

   Going back to equation  (\ref{eq:w3}) and writing it in Fourier
space  we obtain  for the power spectrum
\begin{equation}
P(k,t)=\lsb \sum_{n=0}^{\infty} \frac{1}{n!}\lb \frac{-b^2 k^2 <u^2>}{3} \rb^n \rsb b^2 P_1 (k) \,, \label{eq:w6}
\end{equation}
Summing up  the terms in the square brackets we have
\begin{equation}
P(k,t)= {\rm exp}\lb  \frac{-b^2 k^2 <u^2>}{3} \rb b^2 P_1 (k) \,.
\label{eq:w7} 
\end{equation}
which in real space gives us
\begin{equation}
\xi(x,t)=\frac{1}{(\sqrt{ \pi} 2 L(t))^3} \ini {\rm exp} \lsb-
\frac{(x-x^{'})^2}{4 L(t))^2} \rsb \xi^1 (x^{'},t) d^3 x^{'}  \,, 
\label{eq:w8}
\end{equation}
where 
\begin{equation}
L^2 (t)=\frac{1}{3} b^2 (t) <u^2> \,.
\end{equation}
The length scale $L(t)$ is  the r.m.s. deviation of the particles from
their Lagrangian (or initial) positions at any time $t$. 
 We see that the nonlinear evolution of the two point correlation
function  at large  $x$ corresponds to a convolution of the linear two
point correlation  with a Gaussian whose width is proportional to
$L(t)$. This is 
consistent with our interpretation of the evolution in terms of a
diffusion process. 

For the case when the initial power spectrum has the form 
\begin{equation}
P_1(k)=A e^{- \alpha^2 k^2} k^n \,, \label{eq:w9}
\end{equation}
 using equation (\ref{eq:w3}) at small $k$, we have for the nonlinear power spectrum at small $k$  
\begin{equation}
P_1(k)=A e^{-(\alpha^2+L^2(t)) k^2} k^n \,,\label{eq:w10}\,.
\end{equation}
Using equation (\ref{eq:w9}) and (\ref{eq:w10}), and using the fact that   
\begin{equation}
\int e^{i k x} e^{-\beta^2 \alpha^2 k^2} P_1(k) d^3
k=\frac{1}{\beta^{3+n}} \int_{-\infty}^{\infty}  e^{i k
\frac{x}{\beta}} e^{- \alpha^2 k^2} P_1(k) d^3 k 
\end{equation}
we obtain for the nonlinear two point correlation function at large $x$
\begin{equation}
\xi_1(x,t)=\lsb 1+ \lb \frac{L(t)}{\alpha} \rb ^2
\rsb^{-\frac{3+n}{2}} \xi^{(1)}_2(\frac{x}{\sqrt{1+ \lb
\frac{L(t)}{\alpha} \rb ^2 }},t)\,. \label{eq:w11}
\end{equation} 

This formula relates the nonlinear two point correlation at some
separation $x$ at a time t to the linear two point correlation at a
smaller separation at the same time. Thus, at large $x$, for small
values of the two point correlation,  we have information
being transferred out from the smaller scales to the larger scales. 

We next numerically investigate the evolution of the two point
correlation function at large separations for the initial power
spectrum $P_1(k)=.5 e^{-k^2} k$. Figure 1 shows the function $\xiox$
as  a function of x. This function multiplied by the square of the
scale factor gives the linear two point correlation $\xioxt$. 
At large $x$ the function $\xiox$ has a negative sign and 
a power law behaviour $x^{-4}$. We investigate the evolution of the
two point correlation function at the large separation $x=10$. We do
this using four different approximations which we list below: \\
(a). linear perturbation theory \\
(b). linear theory + the lowest order nonlinear correction using GD
(paper II). \\
(c). the result obtained from summing the whole perturbation
series for the ZA with the extra assumptions about the evolution at
large separations made in this section i.e. equation ({\ref{eq:w11})\\
(d). the non-perturbative two point correlation calculated using ZA
(\ref{eq:a9}) 

This exercise allows us to investigate two different issues. The
first thing that we can check is how well ZA approximates GD. This can be
done by comparing 
(b) with (c) and (d). In this section we have made some assumptions
about the large $x$ behaviour of the two point ocrrelation function
and arrived at the diffusion picture for the evolution. We can put
these assumptions to test by comparing (c) with (d).
  The results are shown in figure 2. We find all the four approximations
match in the early stages of the evolution. The two point
correlation at this separation is initially negative and this value
evolves according to linear theory where it gets multiplied by
$b^2$. The different approximations start to differ
as the evolution proceeds.  The first thing to note is that they start
to differ much before $\xi(x,t) \sim 1$ when one would naively expect
the perturbation  series to break down.  This is a consequence of the
non-local nature of the nonlinear terms for the two point correlation.
As discussed in paper II, this can be understood using equation
{(\ref{eq:w2}) 
\begin{eqnarray}
<u^2>= \ini \xiox x dx \nonumber
\end{eqnarray}
which shows that the nonlinear correction depends on the linear two
point correlation 
condition at all the scales and the major contribution to this
integral comes from the small scales. The small scales become strongly
nonlinear very early in the evolution  and it is because of this that the
nonlinear term starts contributing at large $x$ even when $\xi(x,t)
\ll 1$. In all 
the approximations (i.e. (b),(c) and (d)) the effect of the initial deviation
from the linear theory is to make the growth rate faster than $b^2(t)$. 
In the initial stages approximations (b), (c) and (d) exhibit
qualitatively similar behaviour  but as 
the evolution proceeds we find that (d) starts showing a
behaviour completely
different from (b) and (c). We find that the rapidly decreasing function (d) 
slows down its  decrease and then starts to increase. This is quite
different from the behaviour of (b) and (c) which continue to decrease. This
difference is because of the effects of multi-streaming. In ZA the
correlations get washed out after multi streaming occurs.
 Until the onset of
multistreaming the diffusion picture (c) matches quite well with the
full ZA i.e. (d).  A comparison of (b),(c) and (d) shows that ZA
qualitatively predicts the same behaviour as GD and the quantitative
difference may be attributed to the difference in the  diffusion
coefficients. 
	In the case of the actual gravitational dynamics
(non-perturbative) we expect that the results may be different because
there the 
particles will get 'stuck' in bound objects once multistreaming
occurs (e.g. the adhesion model; Gurbatov, Saichev and Shandarin 1989)
As a result of this the mean square 
displacement of the 
particles will be much less than in ZA or in perturbative GD. Although
we expect this diffusion picture to hold for the actual evolution of
the two  point correlation function at large $x$, the perturbative
treatment of  GD and also calculations using ZA may overestimate what
would be obtained in N-body simulations. Incidentally, the regime
treated here would be difficult to study using such simulations since
it involves the low amplitude tail of the two point correlation
function which would be limited by the size of the box and it would
require a large dynamical range.      

\section{The  3 point correlation function.}
We use ZA to follow the evolution of the N point
correlation function. It is possible to do this nonperturbatively by
following a line of reasoning very similar to that in section 3.
However since
ZA  is a good substitute for the gravitational
dynamics only in the weakly nonlinear regime we prefer to carry out the
investigation perturbatively.

We first consider the evolution of the ensemble averaged N point
distribution function $\rho_N(x^a,u^a,t)$. This is a generalization of the
ensemble averaged two point distribution function introduced in section 3
and the superscript $a$ refers to the various points i.e.1,2... N in phase
space which are arguments of this function. Using equation (\ref{eq:a2})
we obtain for the time evolution of this function
\begin{equation}
\rho_{\rm N} (x^a,u^a,t)=\rho_{\rm N} (x^a-b(t) u^a,u^a,t_0) \label{eq:z1}\,.
\end{equation}
Expanding this in a perturbative series and using $a_1,a_2 \,...\, a_n$
for $n$ indices that independently take values between 1 and N, and using
$\mu_1,\mu_2\,... \, \mu_n$ for $n$ corresponding Cartesian components,
we have
\begin{equation}
\rho_{\rm N} (x^a,u^a,t)=\sum_{n=0}^{\infty} \frac{(- b)^n}{n!}
u^{a_1}_{\mu_1} u^{a_2}_{\mu_2}\,...\, u^{a_n}_{\mu_n} \,
\pl^{a_1}_{\mu_1}\pl^{a_2}_{\mu_2}\,...\,\pl^{a_n}_{\mu_n}\, \rho_{\rm N}
(x^a,u^a,t_0)\,. \label{eq:z2}
\end{equation}
For both the kinds of indices the Einstein summation convention holds and
all the $a^{i}$s have to summed over the range 1 to N whenever they appear
twice and the $\mu_i$s have to be summed over the three cartesian components 
whenever the indices are repeated.
  
To calculate the N point correlation function we take velocity moments of
the N  point distribution function

\begin{equation}
\ini \rho_{\rm N}(x^a,u^a,t) d^{3N} u=\sum_{n=0}^{\infty} \frac{(- b)^n}{n!}
\pl^{a_1}_{\mu_1}\pl^{a_2}_{\mu_2}\,...\,\pl^{a_n}_{\mu_n}
 < u^{a_1}_{\mu_1} u^{a_2}_{\mu_2}\,...\, u^{a_n}_{\mu_n}> \,.
\end{equation}
All the terms where $n$ is odd are zero and only the terms with even $n$
contribute. We also have
\begin{equation}
 < u^{a_1}_{\mu_1} u^{a_2}_{\mu_2}\,...\, u^{a_{n}}_{\mu_{n}}
 u^{b_1}_{\nu_1} u^{b_2}_{\nu_2}\,...\, u^{b_{n}}_{\nu_{n}} >=
 <u^{a_1}_{\mu_1} u^{b_1}_{\nu_1>}\,...\, <u^{a_{n}}_{\mu_{n}}
u^{b_{n}}_{\nu_{n}} > + {\rm permutations} \,.
\end{equation} Using the fact that $\pl^{a_1}_{\mu_1}
\pl^{a_2}_{\mu_2} \,...\,
\pl^{a_{2n}}_{\mu_{2n}}$ is symmetric in all the indices, we can add
up all the permutations to obtain for the terms with even $n$
\begin{eqnarray}
\pl^{a_1}_{\mu_1} \, \pl^{b_1}_{\nu_1} \,...\, \pl^{a_n}_{\mu_n} \,
\pl^{b_n}_{\nu_n} & & < u^{a_1}_{\mu_1} \, u^{b_1}_{\nu_1} \,...\,
u^{a_n}_{\mu_n} \, u^{b_n}_{\nu_n}> = \\ \nonumber & &
\frac{(2n)!}{2^n n!}
\pl^{a_1}_{\mu_1} \, \pl^{b_1}_{\nu_1} \,...\, \pl^{a_n}_{\mu_n} \,
\pl^{b_n}_{\nu_n} \lsb T^{a_1 b_1}_{\mu_1 \nu_1} \,...\, T^{a_n
b_n}_{\mu_n \nu_n} \rsb \,.
\end{eqnarray} where $T^{a b}_{\mu \nu} =<u^a_{\mu} u^b_{\nu}>$ is the
covariance matrix introduced in section 3 generalized for the N point
distribution function.

 Using this in equation (\ref{eq:z2}), we have
\begin{equation}
\ini \rho_{\rm N}(x^a,u^a,t) d^{3N} u = \sum_{n=0}^{\infty}
\frac{(b^2)^n}{2^n n!}
\pl^{a_1}_{\mu_1} \, \pl^{b_1}_{\nu_1} \,...\, \pl^{a_n}_{\mu_n} \,
\pl^{b_n}_{\nu_n} \lsb T^{a_1 b_1}_{\mu_1 \nu_1} \, \, ...\, T^{a_n
b_n}_{\mu_n \nu_n} \rsb\label{eq:z3}
\end{equation}
 In the above equation, for a fixed value of $n$, there
will be a term with n pairs $(a_1 b_1),(a_1 b_2)...(a_n b_n)$ where
each index is independently summed over values 1 to N. Thus, for a
fixed value of $n$, the total contribution is a sum of ${\rm
N}^{2n}$ terms each corresponding to a different set of values for
the position indices. In any one of these ${\rm N}^{2n}$ terms there
can be two kinds of pairs \\ A. if $a_i=b_i$, then $T^{a_i b_i}_{\mu_i
\nu_i}=-\frac{1}{3} \delta_{\mu_i \nu_i} \nabla^2 \phi(0)$ is a
constant \\
 B. if $a_i \neq b_i$ then $T^{a_i b_i}_{\mu_i \nu_i}=
\pl^{a_i}_{\mu_i} \pl^{a_i}_{\nu_i} \phi(a_i,b_i)$ is a function of
the separation between these two points. \\ Any of the terms can be
representd by a directed graph with N vertices and $n$ edges. The
pairs of the kind A correspond to an edge connecting a vertex to
itself and a pair of the kind B corresponds to an edge connecting two
different vertices (figure(3)).  The integral $\ini \rho_{\rm
N}(x^a,u^a,t) d^{3N} 
u$ then corresponds to a sum of graphs with N vertices and the number
of edges going from 0 to infinity.

The quantity $\ini \rho_{\rm N}(x^a,u^a,t) d^{3N} u \, d^3 x^1 d^3 x^2
\,..\, d^3 x^{\rm N}$ is the mean number of particles we expect to
find in the volume $d^3 x^1$ at $x^1$ and $d^3 x^2$ at $x^2$ and
... $d^3 x^{\rm N}$ around $x^{\rm N}$. This has contribution from the
lower (i.e N-1,...1 point) correlation functions also. The residue
when the contributions from the lower correlation functions have been
removed, is called the reduced N point correlation
function. Henceforth we shall refer to the reduced N point correlation
function as the N point correlation function.  The graphs that do not
connect all the N points correspond to functions that do not refer to
all the N points and these are the contributions from the lower
correlations. The reduced N point correlation can be calulated by
considering only the connected graphs with N vertices. The lowest
order contribution to the N point correlation corresponds to the
connected graphs with the least number of edges. These graphs are the
tree graphs and they have N-1 edges. The other terms that contribute
to the N point correlation can be generated by adding more edges to
the tree graphs. 

We use equation (\ref{eq:z3}) to calculate the three point correlation
function. The lowest order at which the three point correlation
develops is $n=2$ and this can be written as

\begin{equation}
\zeta^{(1)} (1,2,3,t)=\frac{b^4}{2} \pl^{\ap_1}_{\mu_1} \,
\pl^{\ap_2}_{\mu_2}\, \pl^{\ap_2}_{\mu_3} \, \pl^{\ap_3}_{\mu_4} \lsb
T^{\ap_1 \ap_2}_{\mu_1 \, \mu_2} T^{\ap_2 \ap_3}_{\mu_3 \mu_4} \rsb
\label{eq:q1}
\end{equation}
 where $\ap_1,\ap_2$ and $\ap_3$ are to be summed over
all possible permutations of $1,2$ and $3$. Equation (\ref{eq:q1})
correponds to the only possible tree graph with three vertices
$\ap_1,\ap_2$ and $\ap_3$, and two edges $(\ap_1,\ap_2)$ and
$(\ap_2,\ap_3)$ (figure(4)).

Using
\begin{equation}
\pl_{\mu} \nabla^2 \phi(x)=\frac{x_{\mu}}{x^3} \ini \xioy y^2 dy=
\frac{1}{3} x_{\mu} \xibox
\end{equation} we have
\begin{eqnarray}
& &\zeta^{(1)}(1,2,3,t)=\frac{b^4}{2} \lsb (1+ \cos^2 \txy ) \xioxt
\xioyt \right. \nonumber \\ &+& \left.
 \cos \txy \frac{2}{3} \frac{d}{dx} \xioxt y \xiboyt 
+\frac{2}{3}(1-3 \cos^2 \txy) \xioxt \xiboyt \right. \nonumber \\ &-& \left.
 \frac{1}{3}(1-3 \cos^2 \txy) \xiboxt \xiboyt \rsb \label{eq:z4}
\end{eqnarray}
 where
\begin{eqnarray} 
x=x^{\ap_2}-x^{\ap_3}\,\,\,\,\,\, ,
\,\,y=x^{\ap_2}-x^{\ap_3} \, \nonumber 
\end{eqnarray}
and
\begin{eqnarray}
\txy=\frac{x_{\mu} y_{\mu}}{xy}
\end{eqnarray}

This explicitly exhibits the dependence of the lowest order induced
three point correlation function on the initial two point correlation
function. We see that the three point correlation depends on both
$\xioxt$ and $\xiboxt$. Thus we see that the small scales can
influence the three point correlation at large scales through the
quantity $\xiboxt$. The lowest order induced three point correlation
function calculated using ZA is very similar to that calculated by
studying gravitational dynamics perturbatively at the lowest order
beyond the linear theory (paper I) and the difference is only
in the numerical factors .  

 We next calculate the higher order terms that
contribute to the three point correlation function. These are
generated by adding more edges to the tree graphs. Figures 5 and 6 
illustrates the simplest cases where we add only one edge to the tree  
graph. Next consider any of the  graphs
with $n>2$ edges. In this graph the tree graph can be embedded in
$C^n_2$ ways. Using this in equation (\ref{eq:z3}) we have
\begin{eqnarray}
\zeta(1,2,3,t)=\sum_{n=0}^{\infty} \frac{b^{2(n+2)}}{2^{n+1} n!}
\pl^{a_1}_{\mu_1} \, \pl^{b_1}_{\nu_1} \,.&.&.\, \pl^{a_n}_{\mu_n} \,
\pl^{b_n}_{\nu_n}
\pl^{a^{'}_1}_{\alpha_1} \pl^{a^{'}_2}_{\alpha_2} \pl^{a^{'}_2}_{\alpha_3}
\pl^{a^{'}_4}_{\alpha_4} 
  \lsb T^{a_1 b_1}_{\mu_1 \nu_1} \,..\right. \nonumber \\ & .& \left. ...
.\, T^{a_n b_n}_{\mu_n \nu_n}
T^{a^{'}_1 a^{'}_2}_{\alpha_1 \alpha_2} 
T^{a^{'}_2 a^{'}_3}_{\alpha_3 \alpha_4}  \rsb
\end{eqnarray} 
As discussed in the previous section, at large $x$ the
contributions from the terms with $a_i=b_i$ will dominate. i.e. at the
lowest order, graphs of the kind shown in figure 5 .  Thus, at
large $x$ the three point correlation function may be written as
\begin{equation}
\zeta(1,2,3,t)=\sum_{n=0}^{\infty} \frac{b^{2n}}{2^n n!} (-\frac{1}{3}
\nabla^2 \phi(0))^n (\nabla^{a_1})^2 (\nabla^{a_1})^2 \,..\,
(\nabla^{a_n})^2 \xi^{(1)}_3(1,2,3,t)
\end{equation}
 where the index $a_i$ indicates at which point the
Laplacian acts, and it is to be summed over the values $1,2$ and $3$.
In Fourier space we have
\begin{equation}
 F_3(k^1,k^2,k^3,t)=\sum_{n=0}^{\infty}
\frac{b^{2n}}{2^n n!} (\frac{1}{3} \nabla^2 \phi(0))^n
((k^{a_1})^2+(k^{a_2})^2 \,...\,(k^{a_n})^2) F^{(1)}_3(k^1,k^2,k^3,t)
\end{equation}
 where $F_3$ is the Fourier transform of the three point
correlation and $F^{(1)}_3$ is the Fourier transform of the lowest
order three point correlation.  The terms can be summed up to obtain
\begin{equation}
 F_3(k^1,k^2,k^3,t)={\rm exp}\lsb-\frac{1}{2} \frac{b^2 <u^2>}{3}
((k^1)^2 +(k^2)^2 +(k^3)^2)\rsb F^{(1)}_3(k^1,k^2,k^3,t) \,
\end{equation}
which gives us in real space
\begin{equation}
\zeta(x^1,x^2,x^3,t)=\frac{1}{(\sqrt{2 \pi}  L(t))^9} \int {\rm exp}
\lsb- \frac{(x^a-y^a)^2}{2 L^2(t)} \rsb \zeta^{(l)} (y^1,y^2,y^3,t) d^9
y \,. 
\end{equation}
Thus, at large separation, the effect of including the higher
order terms for the three point correlation function is to convolve
the lowest order induced three point correlation with a Gaussian of
width $L(t)$. As with the two point correlation function, this too can be
interpreted in terms of a diffusion process. 

\section{Discussion and Conclusions.}
 We find that when we calculate the two point 
correlation function as a series in powers of the growing mode, we get
the same answer if we do the calculation using distribution functions
or if we do it in the single stream approximation. Since the first
method is valid even after multi streaming occurs and the second
method breaks down once multistreaming occurs, once multi streaming
has occured we would expect to get different answers using the two
different methods. But the two results match to all orders in the
expansion parameter. We therefore conclude that even though these
equations are valid in the multi streamed epoch, if we start from 
single streamed initial conditions we cannot perturbatively calculate
any effect due to multistreaming e.g. vorticity, pressure.
This limitation arises from the fact that the full two point
correlation function for ZA, which includes the effects of
multistreaming, is an exponential in $\frac{1}{b^2}$. All the
derivatives of the function $\frac{1}{b} e^{-\frac{A}{b^2}}$  vanish
at $b=0$. As a result, if we try to expand this
function in a series in powers of $b$ around $b=0$, we find that
coefficients of all the powers of $b$ are zero. 
If one considers the power spectrum instead,  it is of the form
$e^{-\alpha k^2 b^2}$.  This 
function can be expressed as a power series in $b^2$ and one might
that it is possible to perturbatively study  the effects of
multi-streaming by working in 
Fourier space instead of real space. Such a conclusion would be
erronous as none of the terms in this expansion would have the
effects of multi-streaming. It would be possible to study the effects
of multi-streaming only if it were possible to sum the whole series.
This point is further illustrated in an appendix where we consider a
simpler example where a similar situation occurs.

Shandarin and Zel'dovich (1989) present a formula for $N$, the mean
number of streams at 
any point, in a situation where the particles are moving in one
dimension under ZA. At small $b$ this formula is of the form
$N=1+e^{\frac{-A}{b^2}}$ where $A$ is a constant characterizing the initial
conditions. If we expand this in powers of $b$, the coefficients for
all the terms are zero and  we find  that the mean number of streams
is one. This confirms that the effects of multi streaming cannot
be studied perturbatively. Although in this analysis we used ZA, we
expect this to hold for the full gravitational dynamics too, as
derived at the lowest order of nonlinearity in paper II. 

In our comparison of the two point correlation function at large
separations we find that the results obtained using ZA are quite
similar to the lowest order nonlinear results obtained using GD and
both of them can be interpreted in terms of a diffusion process where
the rearrangement of matter on small scales affects the two point
correlation at large scales. In ZA, for an initial power
spectrum with $n>-1$, the mean square displacement of the
particles from their original positions is $L^2(t)=b^2(t) <u^2>$ and
this makes its appearance in the formula for the nonlinear corrections
to the two point correlation function obtained using ZA. Interpreting
the results from GD in a similar fashion, for an initial power spectrum
with $n>0$, we have $L^2(t) \sim .58 b^2(t) <u^2>$. In paper II we
also considered the case with $n=0$ and for this case we found $L^2(t)
\sim 1.49 b^2(t) <u^2>$. The differences can be understood in terms of
the fact that in ZA the particles move along trajectories calculated
using linear GD, whereas when we take into account nonlinear
corrections, the trajectories get modified by the tidal forces. In the
equations for the evolution of the two point correlation function the
tidal force acts through the three point correlation function. The
tidal force of the third particle (in the three point correlation), will
cause the other two particles to move towards or away from one
another. This effect will be strongly dependent on the spatial
behaviour of the three point correlation function. For the cases with
$n>0$ the induced three point correlation has the hierarchical form at
large $x$ whereas s for the case with $n=0$ the induced three point
correlation does not have this form. We propose that it is because of
this that the effect of the tidal forces is different in these two
cases and in the former case the effect of the tidal forces is to
reduce the mean square diplacements relative to ZA whereas in the
latter case it increases it. Thus indirectly, it is a diagnostic of the
effect of the backreaction of the three point correlation function on
the pair velocity which in turn effects the two point correlation.

We find that for ZA, at large $x$, we can sum up all terms in the
perturbative series and the nonlinear two point correlation function
is related to the linear two point correlation by a convolution with a
Gaussian of width $\propto L(t)$.
We also find that for special initial conditions where the power
spectrum has a Gaussian cutoff at large $k$, the evolution at large
$x$ can be described by a simple scaling relation according to which
the information propagates outward.         
 
We also find that this picture based on diffusion gives a good
description of the evolution under ZA until the onset of
multistreaming.  Based on this we suggest that the evolution of 
of the two point correlation function in GD can also be
described by a diffusion process until the onset of multistreaming.

We have calculated the lowest order induced three point correlation
function using ZA and we find that it is very similar to the result
obtained using GD and the two differ only in the numerical factors. 
We also investigate the effect of the higher order nonlinear terms and
we find that at large $x$ we can sum the whole perturbation series.
We find that the expression obtained after taking into account the
nonlinear corrections is related to to the lowest order three point
correlation function by a convolution with a Gaussian of width
$\propto L(t)$. This is very similar to the evolution of the two
point correlation function at large separations.

It can be shown that a similar relation holds for the higher
correlation functions also but we do not pursue this matter in this
paper.

\acknowledgements  The author would like to thank  Rajaram
Nityananda for his advice, encouragement, and for many 
very useful suggestions and discussions. 

\appendix
\section{Appendix}
Consider a Gaussian function of the variable $x$ with standard deviation 
$\sigma$.  We are interested in the power series expansion of this
function in $\sigma$ around
$\sigma=0$. We can do this expansion by taking the Fourier transform
of the Gaussian i.e. 
\begin{equation}
\frac{1}{\sqrt{2 \pi} \sigma} e^{-\frac{x^2}{2 \sigma^2}} = \frac{1}{2
\pi} \int e^{i k x} e^{-\frac{1}{2} k^2 \sigma^2} dk
\end{equation}
and then doing a Taylor expansion (convergent) of $e^{-\frac{1}{2} k^2
\sigma^2}$.  We then get 
\begin{equation}
\frac{1}{\sqrt{2 \pi} \sigma} e^{-\frac{x^2}{2 \sigma^2}} = \frac{1}{2
\pi} \int e^{i
k x} \sum_{n=0}^{\infty}\lb -\frac{1}{2} k^2 \sigma^2 \rb^n
\frac{1}{n!} dk \label{eq:ap2}
\end{equation}
which gives us 
\begin{equation}
\frac{1}{\sqrt{2 \pi} \sigma} e^{-\frac{x^2}{2 \sigma^2}} = 
 \sum_{n=0}^{\infty} \frac{1}{n!} \lb \frac{1}{2}  \sigma^2
\frac{d^2}{dx^2} \rb^n \delta(x) \,. \label{eq:ap3}
\end{equation}

Equation (\ref{eq:ap3}) can also be derived if we take the Gaussian
function and directly do a Taylor expansion in $\sigma$ i.e. without
going to Fourier space. 

We see that the series expansion is entirely made up of Dirac delta
functions and its derivatives and hence it has nonzero value only when
$x=0$. This should be compared with the original Gaussian function which has
nonzero value even if $x \neq 0$. We see that in this case the
Taylor  expansion fails to capture an important aspect of the original
function and we can attribute this to the fact that we are doing the 
Taylor expansion of a function which is an exponential in
$\frac{1}{\sigma^2}$.

 If instead of working in real space we work in Fourier space, we  find
that  we have to deal with the
Taylor  expansion of a function which is an exponential in $\sigma^2$
instead of $\frac{1}{\sigma^2}$.
There is no problem in expanding this function in a Taylor series and
on might be led to think that the limitation of the Taylor expansion in   
real space can be overcome by going to Fourier space. But this turns
out to be wrong.  On comparing equations (\ref{eq:ap2}) and
(\ref{eq:ap3}) we see that each term in the expansion in Fourier
space corresponds to some derivatives of a Dirac delta function and
hence it cannot capture any of the effects missed out if the analysis
is done in real space. These effects can be included only if one is
able to sum the series in Fourier space and then do the Fourier
transform.

\newpage
\begin{figure}
\plotfiddle{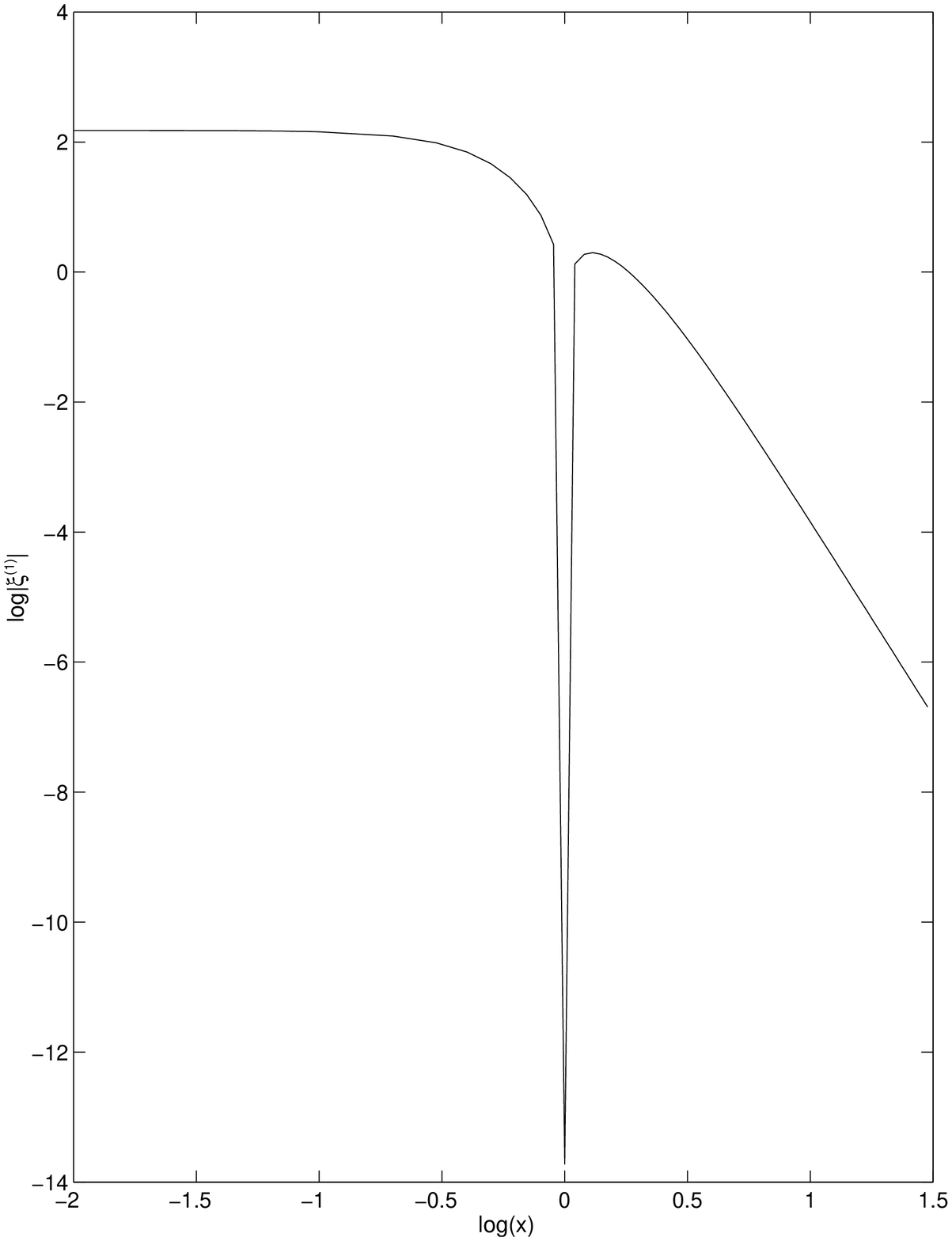}{4in}{0}{60}{60}{-200}{0}
\caption{ The initial two point correlation as a function of the
separation for the  power
spectrum $P(k)=.5 e^{-k^2} k$.} \end{figure}

\begin{figure}
\plotfiddle{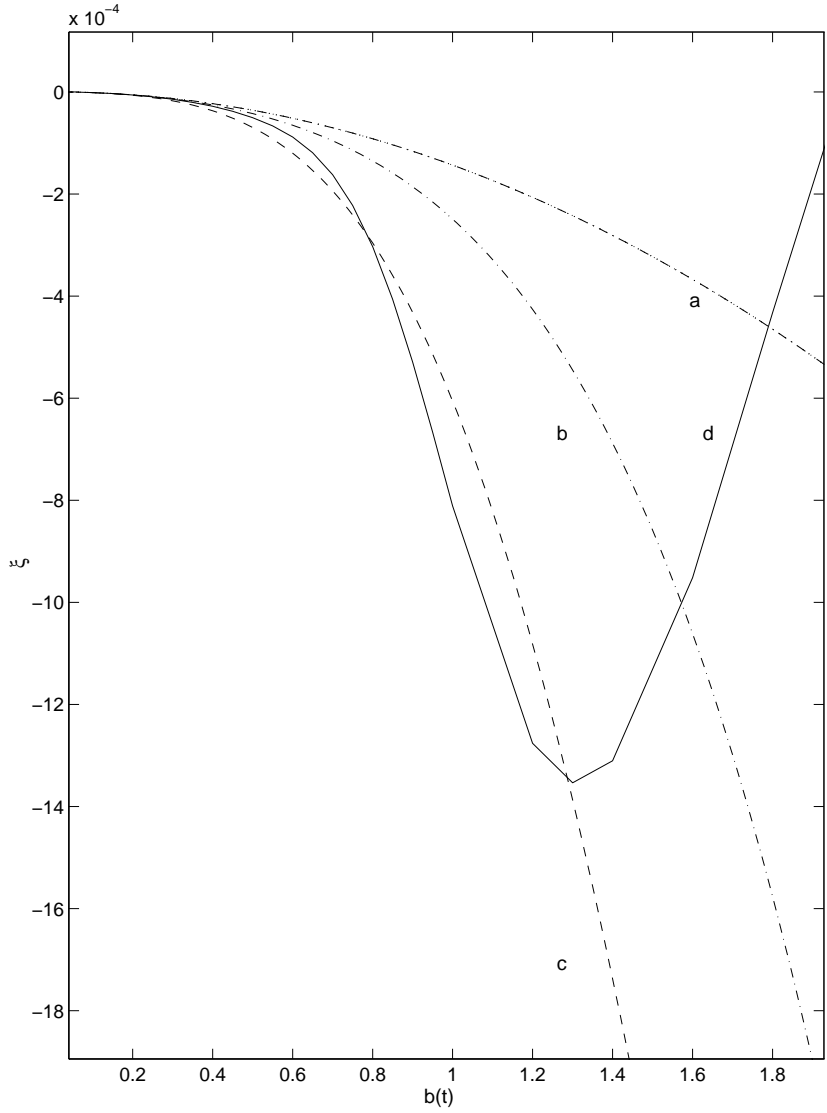}{4in}{0}{60}{60}{-200}{-36}
\caption{ The two point correlation at a fixed separation $x=10$ as a
function of the growing mode $b(t)$ for (a) linear theory, (b) linear
theory + lowest order nonlinear correction using GD (c) nonlinear
evolution using ZA and the assumptions made in section 5 about the
large $x$ behaviour, and (d) nonperturbative ZA} \end{figure}

\begin{figure}
\plotfiddle{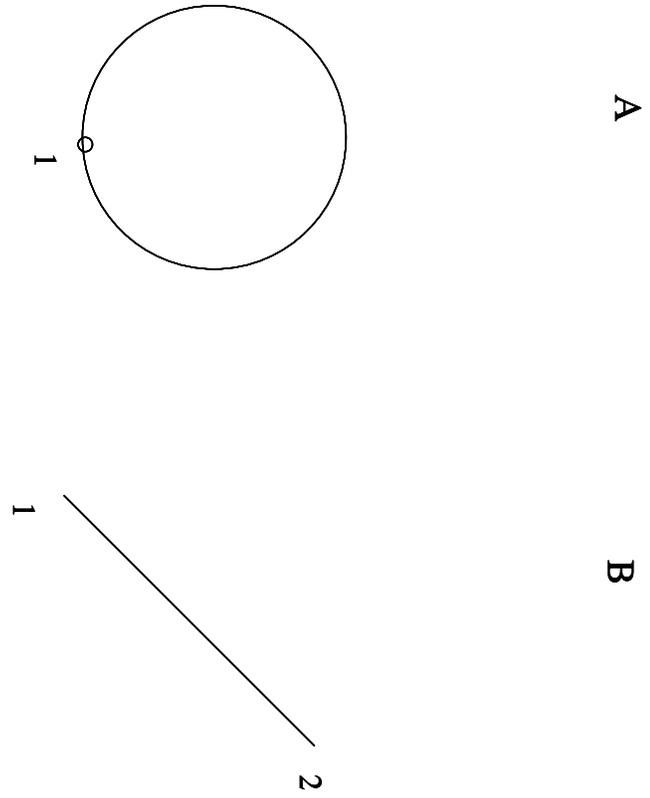}{4in}{0}{60}{60}{-200}{0}
\caption{This shows the two possible kinds of edges A. connects a
vertex to itself B. connects two different vertices.}
\end{figure}

\begin{figure}
\plotfiddle{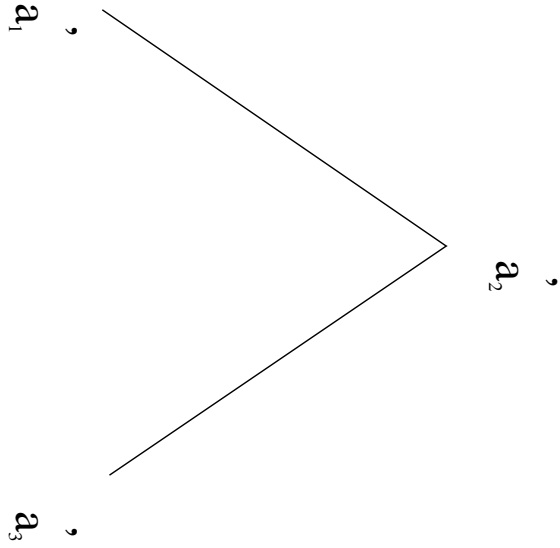}{4in}{0}{60}{60}{-200}{0}
\caption{This shows the tree graph corresponding to the lowest order
induced three point correlation function.}
\end{figure}

\begin{figure}
\plotfiddle{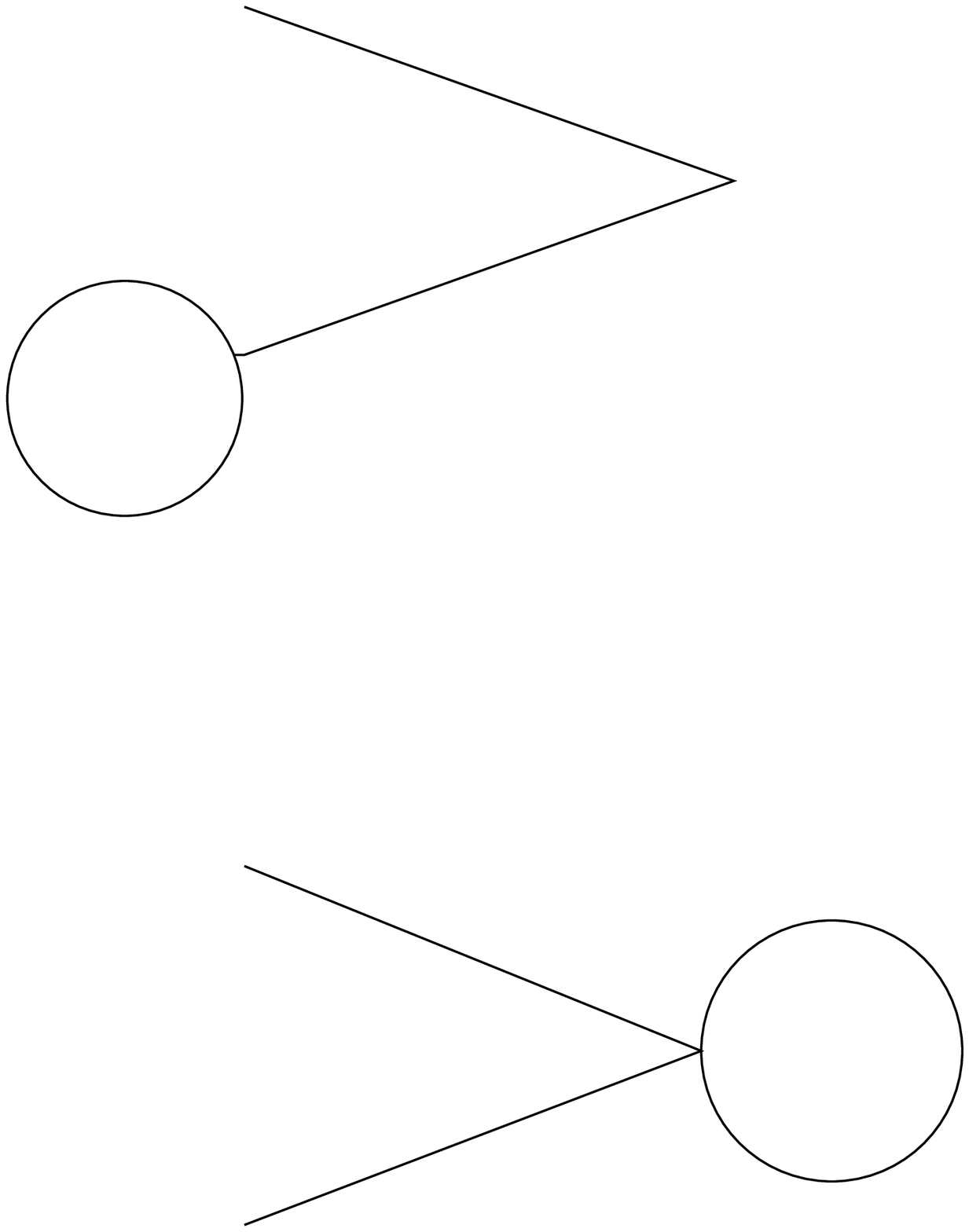}{4in}{0}{60}{60}{-200}{-72}
\caption{This shows some of the graphs corresponding to the 
contribution to the three point correlation function at one order
beyond the lowest. These graphs are all obtained by adding edges to
the tree graph. These graphs show those cases where the extra edge
 connects a vertex to itself. }
\end{figure}

\begin{figure}
\plotfiddle{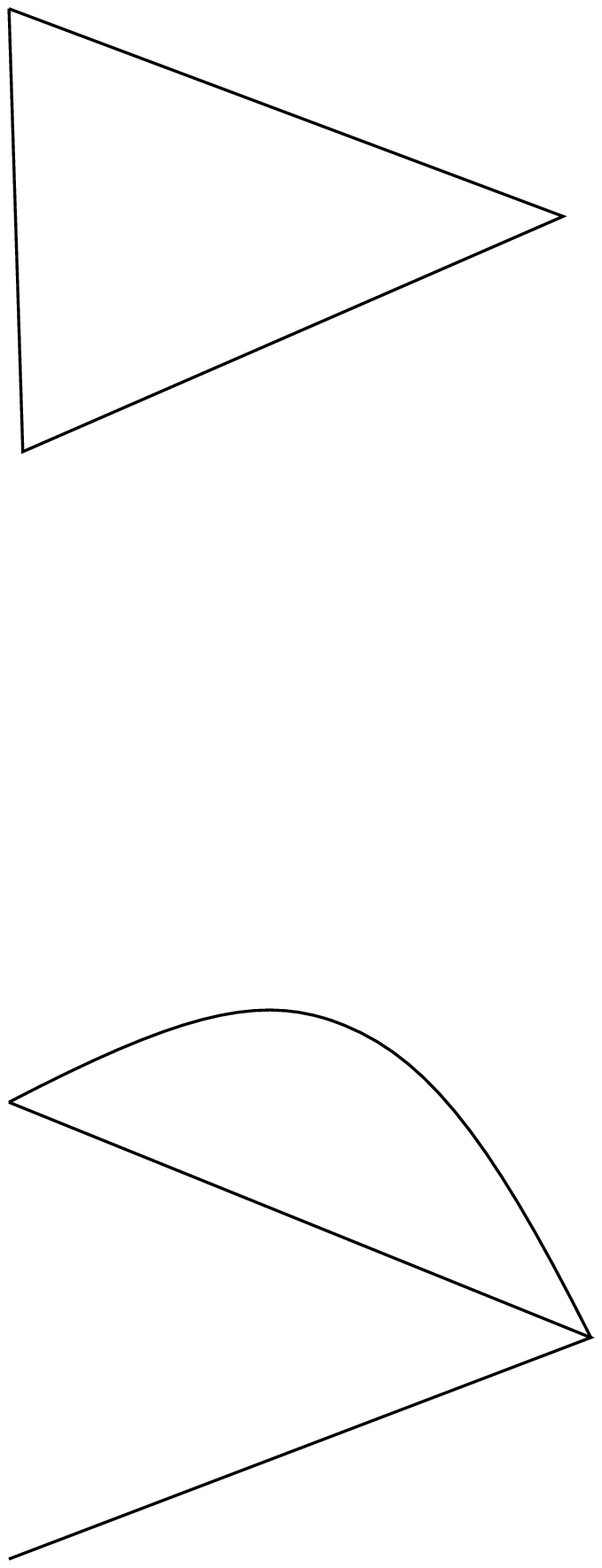}{4in}{0}{60}{60}{-200}{-72}
\caption{This shows some of the graphs corresponding to the 
contribution to the three point correlation function at one order
beyond the lowest. These graphs are all obtained by adding edges to
the tree graph. Thee graphs show those cases where the extra edge
connects two different vertices.}
\end{figure}

\end{document}